\documentclass[sigconf]{acmart}
\acmSubmissionID{2269}
\usepackage[utf8]{inputenc} 
\usepackage[T1]{fontenc}    
\usepackage{hyperref}       
\usepackage{url}            
\usepackage{booktabs}       
\usepackage{amsfonts}       
\usepackage{nicefrac}       
\usepackage{microtype}      
\usepackage{wrapfig,lipsum,booktabs}
\usepackage{multirow}
\usepackage{hyperref}

\usepackage{math_commands}
\usepackage{hyperref}
\usepackage{url}
\usepackage{multirow}
\usepackage{graphicx}
\usepackage{booktabs}
\usepackage{caption}
\usepackage{graphicx}
\usepackage{placeins}
\usepackage{enumitem}
\setlist[itemize]{align=parleft,left=0pt..1em}
\setlist[enumerate]{align=parleft,left=0pt..1em}
\usepackage{algorithm} 
\usepackage{algpseudocode}  
\usepackage{amsthm}
\usepackage{enumitem}
\usepackage{bm}
\newcommand{\eat}[1]{}

\theoremstyle{definition}

\usepackage{subcaption}
\usepackage{xspace}

\mathchardef\mhyphen="2D

\newcommand{\method}{SCCDR\xspace}
\newcommand{\best}[1]{\textbf{#1}}
\newcommand{\xhdr}[1]{\noindent{{\bf #1.}}}

\usepackage{dsfont}

\usepackage{pifont}
\usepackage{ulem}
\makeatletter

\newcommand\addauthornote[1]{%
  \if@ACM@anonymous\else
    \g@addto@macro\addresses{\@addauthornotemark{#1}}%
  \fi}

\newcommand\@addauthornotemark[1]{\let\@tmpcnta\c@footnote
   \setcounter{footnote}{#1}\addtocounter{footnote}{-1}
    \g@addto@macro\@currentauthors{\footnotemark\relax\let\c@footnote\@tmpcnta}}

\makeatother

\usepackage{adjustbox}
\usepackage{float}
\usepackage{booktabs}

\AtBeginDocument{%
  }

\copyrightyear{2025}
\acmYear{2025}
\setcopyright{acmlicensed}\acmConference[WWW Companion '25]{Companion
Proceedings of the ACM Web Conference 2025}{April 28-May 2, 2025}{Sydney,
NSW, Australia}
\acmBooktitle{Companion Proceedings of the ACM Web Conference 2025 (WWW
Companion '25), April 28-May 2, 2025, Sydney, NSW, Australia}
\acmDOI{10.1145/3701716.3715260}
\acmISBN{979-8-4007-1331-6/2025/04}

\begin{document}

\title{Separated Contrastive Learning for Matching in Cross-domain Recommendation with Curriculum Scheduling}

\settopmatter{authorsperrow=4}
\author{Heng Chang}
\orcid{0000-0002-4978-8041}
\email{changh.heng@gmail.com}
\affiliation{%
  \institution{Huawei Technologies Co., Ltd.}
  \country{Beijing, China}
}

\author{Liang Gu}
\authornote{Corresponding authors.}
\orcid{0009-0004-9443-4500}
\email{guliang3@huawei.com}
\affiliation{%
  \institution{Huawei Technologies Co., Ltd.}
  \country{Shenzhen, China}
}

\author{Cheng Hu}
\orcid{0000-0002-0942-3224}
\email{hucheng9@huawei.com}
\affiliation{%
  \institution{Huawei Technologies Co., Ltd.}
  \country{Shenzhen, China}
}

\author{Zhinan Zhang}
\orcid{0009-0008-9693-7661}
\email{zhangzhinan5@huawei.com}
\affiliation{%
  \institution{Huawei Technologies Co., Ltd.}
  \country{Shenzhen, China}
}

\author{Hong Zhu}
\orcid{0000-0003-2943-7997}
\email{zhuhong8@huawei.com}
\affiliation{%
  \institution{Huawei Technologies Co., Ltd.}
  \country{Shenzhen, China}
}

\author{Yuhui Xu}
\addauthornote{1}
\orcid{0000-0002-7109-7140}
\email{xyh6666@gmail.com}
\affiliation{%
  \institution{Huawei Technologies Co., Ltd.}
  \country{Shenzhen, China}
}

\author{Yuan Fang}
\orcid{0009-0000-7978-225X}
\email{frank.fy@huawei.com}
\affiliation{%
  \institution{Huawei Technologies Co., Ltd.}
  \country{Shenzhen, China}
}

\author{Zhen Chen}
\orcid{0000-0001-7997-9888}
\email{zzz.chen@huawei.com}
\affiliation{%
  \institution{Huawei Technologies Co., Ltd.}
  \country{Shenzhen, China}
}

\renewcommand{\shortauthors}{Heng Chang et al.}

\begin{abstract}
Cross-domain recommendation (CDR) is a task that aims to improve the recommendation performance in a target domain by leveraging the information from source domains. Contrastive learning methods have been widely adopted among intra-domain (intra-CL) and inter-domain (inter-CL) users/items for their representation learning and knowledge transfer during the matching stage of CDR. However, we observe that directly employing contrastive learning on mixed-up intra-CL and inter-CL tasks ignores the difficulty of learning from inter-domain over learning from intra-domain, and thus could cause severe training instability. Therefore, this instability deteriorates the representation learning process and hurts the quality of generated embeddings. To this end, we propose a novel framework named SCCDR built up on a separated intra-CL and inter-CL paradigm and a stop-gradient operation to handle the drawback. Specifically, SCCDR comprises two specialized curriculum stages: intra-inter separation and inter-domain curriculum scheduling. The former stage explicitly uses two distinct contrastive views for the intra-CL task in the source and target domains, respectively. Meanwhile, the latter stage deliberately tackles the inter-CL tasks with a curriculum scheduling strategy that derives effective curricula by accounting for the difficulty of negative samples anchored by overlapping users. Empirical experiments on various open-source datasets and an offline proprietary industrial dataset extracted from a real-world recommender system, and an online A/B test verify that SCCDR achieves state-of-the-art performance over multiple baselines.
\end{abstract}



\maketitle

\section{Introduction}\label{sec:intro}
Recommender systems are widely used to provide personalized and relevant suggestions to users based on their preferences and behaviors~\cite{covington2016deep,man2017cross,zhu2019dtcdr,cui2020herograph,xie2020internal,wu2023curriculum}. 
Most real-world large-scale recommender systems employ the conventional two-stage architecture consisting of \textit{matching} and \textit{ranking}. The \textit{matching} module \cite{xie2020internal} (also known as candidate generation \cite{covington2016deep}) aims to enhance the efficiency and diversity of the system, by retrieving a small subset of (typically hundreds of) item candidates from the million-level large corpora. Subsequently, the \textit{ranking} module assigns the specific ranks of items for the final results.

As the scale of recommender systems expands and the diversity of recommendation contexts broadens, the integration of supplementary data sources, or domains, is essential to enhance content comprehensiveness and diversity. Nonetheless, recommender systems frequently confront the challenges of data sparsity and cold-start phenomena, particularly when insufficient or noisy data plague the target domain.
Cross-domain recommendation (CDR) is a task that aims to address this challenge by leveraging the information from source domains that have rich and high-quality data~\cite{hu2018conet,xie2022contrastive,chen2023cross}. In this way, CDR can improve the recommendation performance in the target domain by transferring the knowledge from the source domains.
During CDR, contrastive learning (CL)~\cite{oord2018representation,you2020graph,chen2021exploring} has been widely adopted for representation learning and knowledge transfer, especially in the matching stage, where the goal is to retrieve the most relevant items for a given user~\cite{xie2022contrastive,cui2020herograph}. CL can be applied to both intra-domain (\textit{intra-CL}) and inter-domain (\textit{inter-CL}) nodes, where intra-domain nodes are the users and items within the same domain, and inter-domain nodes are the users and items across different domains.

\begin{figure}
  \centering
  \includegraphics[width=0.4\textwidth]{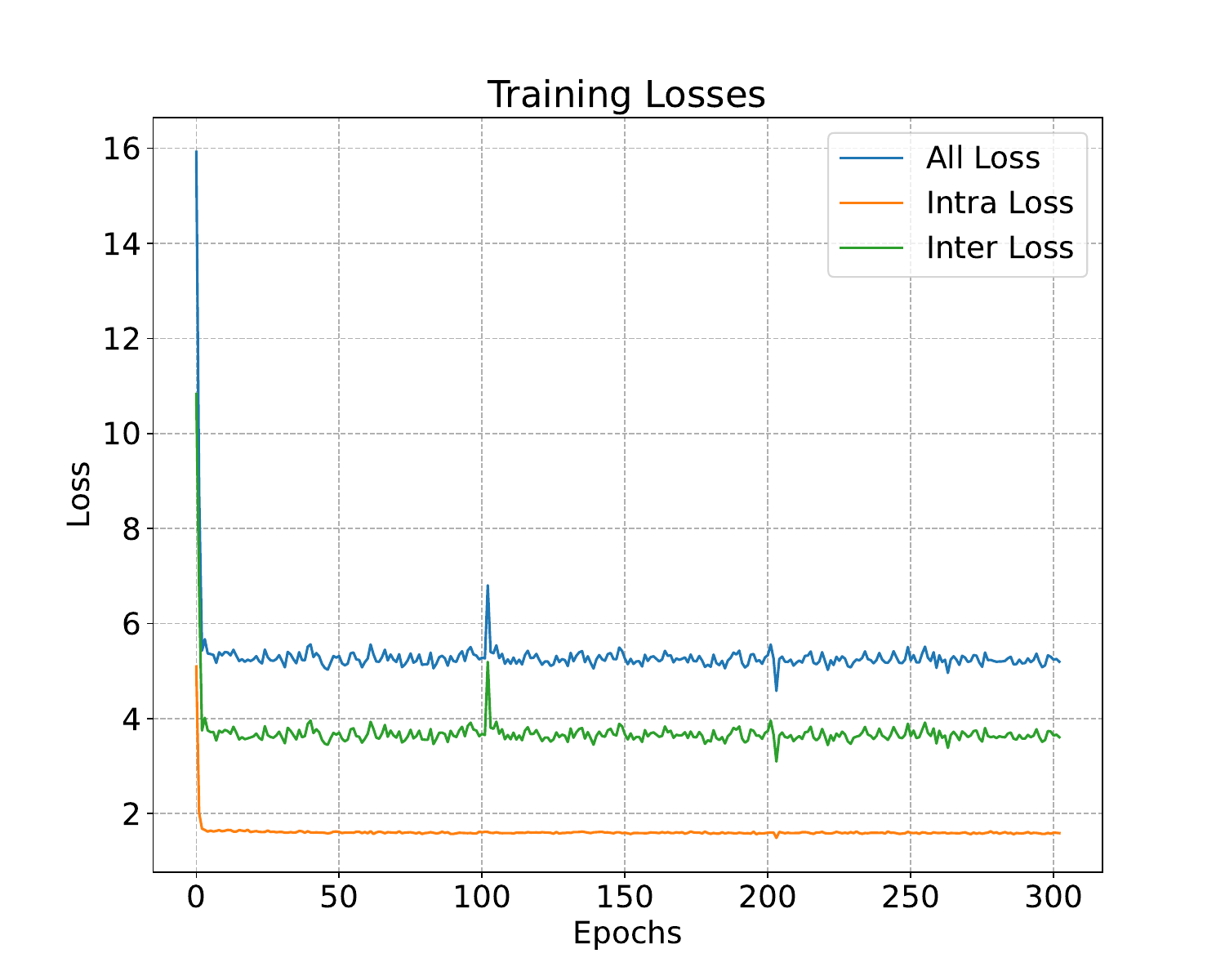}
  \vspace{-5mm}
  \caption{The stability comparison between intra-CL and inter-CL losses during the training of cross-domain recommendation task on Amazon Books-Videos dataset.}
  \Description{The stability comparison between intra-CL and inter-CL losses during the training of cross-domain recommendation task on Amazon Books-Videos dataset.}
  \vspace{-8mm}
  \label{fig:loss_exp}
\end{figure}

\begin{figure*}
  \centering
  \includegraphics[width=0.95\textwidth]{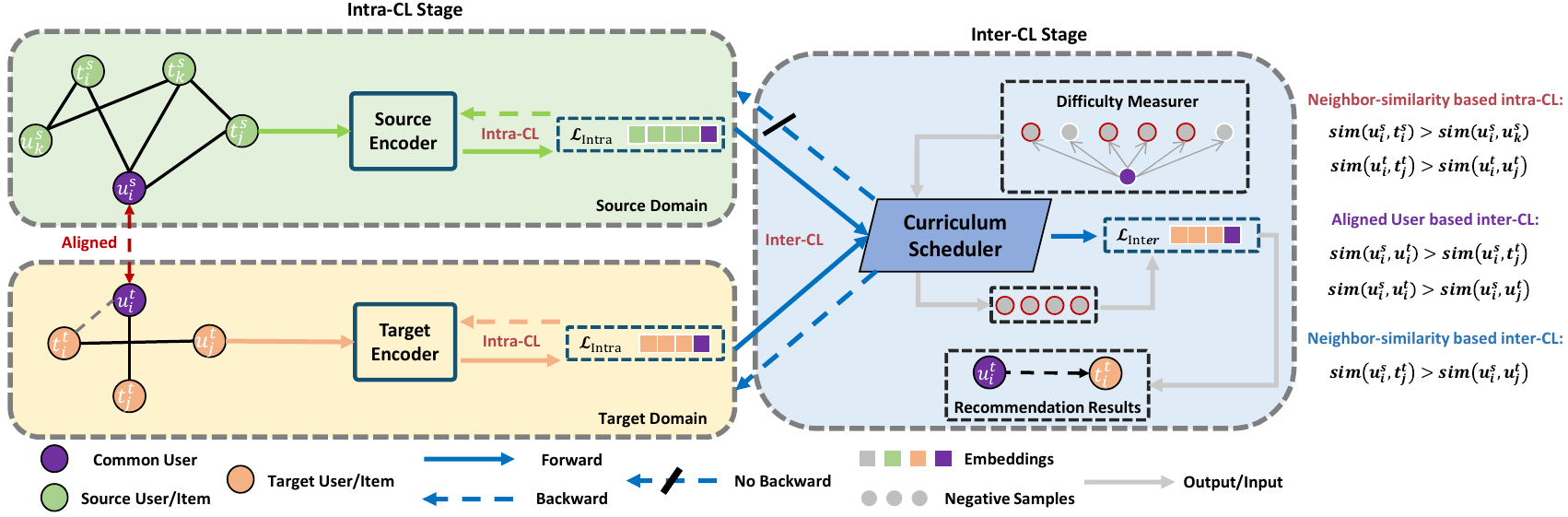}
  \vspace{-4mm}
  \caption{The illustration of our proposed framework \method and the example of the intra-CL and inter-CL losses we considered.} 
  \Description{The illustration of our proposed framework \method and the example of the intra-CL and inter-CL losses we considered.} 
  \vspace{-5mm}
  \label{fig:framework}
\end{figure*}

While the intra-domain is comparably straightforward for the model to learn since the interaction pattern among users and items within the same domain are easier to capture than that in the inter-domain, conventional approaches~\cite{xie2022contrastive,zhang2023collaborative} usually directly employ contrastive learning on mixed-up intra-CL and inter-CL tasks.
However, as shown in Figure~\ref{fig:loss_exp}, we observe that this mixed-up strategy ignores the difficulty of learning from inter-domain over intra-domain, and thus could cause significant training instability (please kindly refer to Section~\ref{sec:method} for more detailed analysis).
This instability, therefore, could hurt the following representation learning process as well as the knowledge transfer from the source domain to the target domain.

To this end, we propose a novel \textit{\textbf{S}}eparated contrastive learning with \textit{\textbf{C}}urriculum Scheduling framework for matching in \textit{\textbf{CDR}}, namely \textit{\textbf{\method}}, to handle this deficiency in training as shown in Figure~\ref{fig:framework}. Specifically, \method consists of two specialized curriculum stages: intra-inter separation and inter-domain curriculum scheduling. 
The former stage explicitly uses two distinct contrastive views for the intra-CL task in the source and target domains, respectively. Meanwhile, the latter stage tackles the inter-CL tasks with a curriculum scheduling strategy that derives effective curricula by accounting for the difficulty of negative samples anchored by overlapping users. Furthermore, a stop-gradient operation is introduced between the two stages to restrict the excessive gradient oriented from the source domain during the process of knowledge transfer.
Our contributions are summarized as follows:
 \begin{itemize}
     \item To the best of our knowledge, we are the first to consider the different roles of intra-CL and inter-CL in training for matching in CDR through a separated intra-CL and inter-CL paradigm with a stop-gradient operation.
     \item To alleviate the impact of the potential noises hidden in the negative samples, we further propose using a curriculum scheduler during the inter-CL stage for better representation calibration.
     \item We conduct extensive experiments on both open-source and proprietary industrial datasets. Empirical results show that our \method framework brings consistent performance improvements over various SOTA baselines regarding both offline and online evaluations, demonstrating its generic and powerful ability for matching tasks in CDR.
 \end{itemize}

\section{Related Work}\label{sec:related-works}
\subsection{Cross-domain Recommendation}
Cross-domain recommendations (CDR) aim to enhance the accuracy of recommendations in a target domain by leveraging knowledge from a source domain~\cite{zhao2023beyond,zhang2023collaborative,zhao2023cross,liu2024graph}. CoNet~\cite{hu2018conet} introduces cross-connections to facilitate dual knowledge transfer across domains, while MiNet~\cite{ouyang2020minet}, featuring inter-level and interest-level attention mechanisms, jointly models users' long-term and short-term interests. DASL~\cite{li2021dual} employs a dual embedding and attention strategy for iterative information transfer between domains. AFT~\cite{hao2021adversarial} employs a generative adversarial network to master feature translations across diverse domains. ~\citeauthor{chen2023cross}~\cite{chen2023cross} consider behavior-level effect during the loss optimization process by proposing a generic behavioral importance-aware optimization framework. Additionally, CDR offers solutions to the cold-start problem, with CCDR~\cite{xie2022contrastive} and SSCDR~\cite{kang2019semi} advocating for contrastive learning and semi-supervised learning approaches, respectively, to compensate for the scarcity of user behavior data.

\subsection{Contrastive Learning in Recommendation}
Contrastive learning (CL), aimed at acquiring high-quality representations through self-supervised techniques, has garnered significant attention across various fields of machine learning~\cite{chen2020simple,zhu2021graph,gutmann2010noise}. Motivated by the accomplishments of CL across different domains, there has been a surge in innovative research efforts that incorporate CL into recommender systems~\cite{qiu2021memory,zhou2020s3,ma2020disentangled}. ~\citeauthor{zhang2022diffusion}~\cite{zhang2022diffusion} propose GDCL to capture the structural properties of the user-item interaction graph more effectively. \citeauthor{yu2022graph}~\cite{yu2022graph} introduce a simple CL method that eschews graph augmentations in favor of injecting uniform noise into the embedding space, thereby generating contrastive views. AdaGCL~\cite{jiang2023adaptive} employs two adaptive contrastive view generators for data augmentation, significantly enhancing the collaborative filtering (CF) paradigm.

\section{Preliminaries}\label{sec:pre}
In this section, we start by defining the notations, then introduce preliminaries of the task of CDR in the matching stage and the basics of contrastive learning with GNN-based encoders. 

Following existing works~\cite{liu2019jscn,xie2022contrastive,wang2021pre}, we model the user-item interactions in the recommendation scenario as graphs. We denote the graph as $\gG = (\gU, \gT, \gE)$, where $\gU, \gT, \gE \subseteq \gU \times \gT$ denote the set of users, items, and edges, respectively. Regarding the CDR problem, we denote a source domain graph as $\gG^{s} = (\gU^{s}, \gT^{s}, \gE^{s})$ and a target domain graph as $\gG^{t} = (\gU^{t}, \gT^{t}, \gE^{t})$. 
The number of nodes for the source and target domains is defined as $N^{s}$ and $N^{t}$.
As individuals, we use ($u_{i}^{s}$, $t_{i}^{s}$) and ($u_{i}^{t}$, $t_{i}^{t}$) to represent the $i\_$th user and item pair in the source domain and target domain, respectively. The overlapped users form a non-empty set $\gS$ such that $\gS \subset \gU^{s}$ and $\gS \subset \gU^{t}$. We use the same subscripts such as $i$ if $u \in \gS$.
Let $\vu_{i}$ and $\vt_{i}$ be the embeddings on the user and item, and $\mathcal{N}(u_i)$ and $\mathcal{N}(t_i)$ be the neighborhood of $i$-th user and item node, respectively.

\subsection{Matching Stage in CDR}
The CDR approach for the matching task seeks to enhance the matching performance of the target domain by leveraging the information from the source domain.
Following \cite{xie2022contrastive}, our focus is on the matching module of the conventional two-stage recommender systems \cite{covington2016deep}. 
The matching module precedes the ranking module and aims to efficiently retrieve a subset of relevant items from a large-scale item pool. The matching module is more concerned with the presence of good items in the retrieved set (often evaluated by the hit rate metric (HIT@N)), rather than the exact order of the items, which is the responsibility of the subsequent ranking module (often assessed by NDCG or AUC)~\cite{lv2019sdm,xie2020internal,chen2023cross}.

\subsection{Graph-based Contrastive Learning in CDR}
Graph-based contrastive learning (GCL) is a novel paradigm for recommender systems that leverages the power of graph neural networks (GNNs)~\cite{gu2020implicit,li2022semi,chang2020restricted,chang2021not,chang2021spectral,chang2022adversarial,wang2023revisiting,xie2023adversarially,wang2024heterophilic,yang2024hyperbolic,xie2024towards,chang2024path} and self-supervised learning~\cite{yu2022graph,jiang2023adaptive,zhang2022diffusion}. GCL aims to learn better user and item representations by maximizing the agreement between different views of the same graph while minimizing the similarity between views of different graphs~\cite{you2020graph}. 
GCL can effectively address the challenges of data sparsity, noise, and heterogeneity in recommender systems, and improve the performance of various downstream tasks.

GCL in CDR usually consists of three main components: graph encoders for source domain $f^{s}$ and target domain $f^{t}$, a view generator $g$, and a contrastive loss $\mathcal{L}$:

\begin{itemize}
    \item The graph encoder is a GNN model that takes a user-item interaction graph ($\gG$) as input and outputs the embeddings of users ($\vu$) and items ($\vt$):
    \begin{equation}
    \vu^{*}, \vt^{*} = f^{*}(\gG^{*}; \mW^{*}),
    \end{equation}
    where $\mW$ denotes network weights, and $*$ indicates either the source or target domain. $f^{s}$ and $f^{t}$ can share parameters (\ie, employed with a single GNN) regarding model design and we choose to distinguish them with two GNNs that are of the same architecture in our framework. A general formula for updating user representation $\vu$ in GNNs is denoted as
    \begin{equation}\label{eq:gnn}
    \vu_i^{(l)} = \sigma \left( \mW^{(l)} \cdot \left(\operatorname{Aggr}(e_{ij}^{(l)}\vu_j^{(l)}, u_j \in \mathcal{N}(u_{i})) + B^{(l)} \cdot \vu_i^{(l)} \right) \right),
    \end{equation}
    where $\vu_i^{(l)}$ denotes the node representation for user $u_{i}$ in the $l$-th hidden layer, $e_{ij}^{(l)}$ is the correlation coefficient between node pair with index $i$ and $j$, $B^{(l)}$ represents the weights for the self-loop of $i$-th node, $\operatorname{Aggr}(\cdot)$ is the function to aggregate neighborhood information, $\operatorname{Comb}(\cdot)$ aims for combining self- and neighbor-information, and $\sigma(\cdot)$ is the activation function. The layer-wise update rule for item embedding $\vt$ is the same as $\vu$ as in Eq.~(\ref{eq:gnn}).
    \item The view generator $g$ for CDR creates different views of users and items from both the source domain and the target domain, such as by adding or removing edges, perturbing node features, or sampling subgraphs. For intra-CL, the views are generated deliberately for the source or target domain and focus on relationships within the single domain such as the neighborhood around the nodes. For inter-CL, the views are produced by considering cross-domain relationships and usually depend on the alignment of overlapped users.
    \item The contrastive loss $\mathcal{L}$ is a function that measures the similarity between embeddings of different views and encourages the embeddings of the same nodes to be closer than the rest of the nodes in the latent space. Representative contrastive losses are BCELoss~\cite{gutmann2010noise}, Triplet loss~\cite{schultz2003learning}, and InfoNCE~\cite{oord2018representation}. Among them, InfoNCE is the most widely adopted application in CDR.
\end{itemize}

\section{Proposed Method: \method}\label{sec:method}
In this work, we propose \method to enhance the cross-domain recommendation in matching via intra-CL and inter-CL separation, and curriculum scheduler in inter-CL alignment.
We begin by introducing the general framework of \method then dive into the details of each module design. 

\subsection{Separated Intra-CL and Inter-CL Paradigm}\label{sec:method1}
Recall the challenge we mentioned in Section~\ref{sec:intro}, we can observe from Figure~\ref{fig:loss_exp} that the naive mixture of the intra-CL and inter-CL tasks could result in unstable training and therefore infect the quality of user and item representations. This also coincides with our intuition that contrastive patterns within the same domain could be easier to capture than those in cross-domain.  The reasons for this are twofold: 1) The proportion of overlapped users is always small compared with the rest of user and item nodes that belong to the same domain. The richer interaction within a single domain makes the representation learning easier from the data perspective.
2) The preference within the same domain is more homogeneous than cross-domain. For example, it is easier to suppose a person prefers classical music from his/her music playlist rather than his watched history of movies.
As a result, the inter-CL knowledge transfer with the help of overlapped users is more difficult unless the contrastive patterns are well-learned in the intra-CL task for individual domains.

Tackling this drawback of the mixture of intra-CL and inter-CL tasks, we propose to distinguish the intra-CL and inter-CL tasks individually. Specifically, we develop a dual-oriented contrastive framework \method that is deliberately designed for the CDR task as in Figure~\ref{fig:framework}. \method explicitly divides the learning of the intra-CL and inter-CL tasks into two stages:
\begin{itemize}
    \item During the \textit{first intra-CL stage}, intra-CL tasks from source and target domains are optimized individually to generate the representation of users $\vu$ and items $\vt$ for each domain.
    \item In the \textit{second inter-CL stage}, \method switches to inter-CL tasks and focuses on the calibration of node embeddings in the target domain with the help of the fruitful source domain interaction.
\end{itemize}

\xhdr{Stop-gradient}
To further stabilize the knowledge transfer and protect the representation of the source domain, we further introduce a stop-gradient operation in the connection between the two stages inspired by the studies on non-contrastive self-supervised learning~\cite{chen2021exploring,caron2021emerging}. 
These non-contrastive approaches utilize a stop-gradient operation to address a dimensional collapse issue that shrinks representations into a reduced-dimensional subspace. This helps to prevent the encoder from outputting the same representations for all inputs and from simplifying the embedding distribution. 

In analogy to them, we adapt stop-gradient during the inter-CL stage and attempt to mitigate the interdependency between user and item embedding distributions, as depicted in the inter-CL stage from Figure~\ref{fig:framework}.
The motivations are also from two perspectives:
1) We assume that the substantial relationships between users and items in the source domain are adequate for learning superior node representations. Therefore, the involvement of the target domain during inter-CL could bring additional noise and then hurt the embeddings from the source domain. Subsequently, this would hinder the learning for nodes in the target domain as well. Therefore, it is desirable to protect the source domain embeddings that are already adequately trained in the intra-CL domain.
2) Meanwhile, the stop-gradient operation is found to be able to break the interdependency between user and item embedding distributions for the recommendation task~\cite{oh2023dual}. Simply combining inter-domain contrasts in a multi-task way leads to breaking the underlying semantic structure of the source domain embeddings. Therefore, including the stop-gradient operation helps to decrease the influence from the target domain by restricting the excessive gradient during knowledge transfer and alleviates the over-aligned source and target distributions. 
We empirically observe that stop-gradients for the source domain ensure individual embedding characteristics (i.e., uniformity), leading to a performance boost (please refer to the ablation study in Section~\ref{sec:ablation}).

\begin{table*}[htbp]
    \centering
    \caption{Statistics of the Amazon and proprietary industrial datasets. We choose 7 domains from Amazon datasets and 2 domains from our collected proprietary industrial dataset to validate the generalization ability of \method.}
    \vspace{-3mm}
    \label{tab:datastatistics}
    \resizebox{0.8\textwidth}{!}{
    \begin{tabular}{lccccccccc}
        \toprule
        & Books- & Books- & Books- & Books-  & Videos- & Cloth- & Kitchen- & Elec- & Industrial: \\ 
        & Videos & Music & Elec & Toys  & Music & Music & Music & Cloth & Music2Videos \\ 
        \midrule
        users          	&344,000   	&6,4920    	&380,874    	&243,265   &51,534     	&4,286      	&8,144       	&142,834   & 7,388,137   \\
        source item    	&222,244  	&119,694   	&220,020   	&205,081  &33,052     	&8,212      	  &22,550        &44,837  & 50,006      \\
        target item     &31,086     &694      &39,171      &8,848    &1,466      &712      &36,018      &21,299    & 94,207  \\
        source edges & 1,826,791  & 643,112   & 1,809,283   & 1,619,884  &287,218  & 15,123     &6,670      &470,057  & 15,313,658  \\
        target edges & 229,147  & 1,251   & 334,807   & 44,484    &3,397  	&1,013     	&60,358      	&123,890    & 7,030,889 \\        
        \bottomrule
    \end{tabular}
    }
\end{table*} 

\subsection{Dual-oriented GCL}
Motivated by the great successes of GNNs, we instantiate each module in the dual-oriented CDR contrastive framework with graph-based contrastive learning as the representative solution.
Note that \method can be easily extended to other contrastive learning approaches by properly designing the graph encoders and view generators.
In the context of GCL, \method considers each component of GCL in CDR as the following:

\xhdr{Graph encoder} GCL can be combined with different GNN architectures by adopting different GNN models, such as GCN~\cite{ICLR2017SemiGCN}, GAT~\cite{velickovic2018gat}, GraphSAGE~\cite{Hamilton2017Inductive}, JK-Net~\cite{Xu2018Representation} or LightGCN~\cite{he2020lightgcn}. 
For the $i$-th user $u_{i}$, we first sample a dynamic sub-neighbor set $\tilde{\mathcal{N}}(u_{i})$ that is randomly generated from ${\mathcal{N}}(u_{i})$ for efficient batch training. The resulting subgraph is denoted as $\tilde{\gG^{s}} = (\tilde{\gU^{s}}, \tilde{\gT^{s}}, \tilde{\gE}^{s})$ for source domain and $\tilde{\gG^{t}} = (\tilde{\gU^{t}}, \tilde{\gT^{t}}, \tilde{\gE}^{t})$ for target domain. We use a user node as the example here and the notions also hold for item nodes. We also denote the formed batch as $\gB$.

For the sake of the efficiency of fast embedding-based retrieval in matching, the calculation of user-item interaction should not be complicated in recommendation tasks. Considering simplicity, we choose to use GraphSAGE~\cite{Hamilton2017Inductive} with JK-Net~\cite{Xu2018Representation} as the same graph encoder on the sampled subgraphs $\tilde{\gG^{s}}$ and $\tilde{\gG^{t}}$ for both source and target domains, respectively:
\begin{align}
    \vu_i^{(l)} & =\sigma \left( \mW^{(l)} \cdot \left(\operatorname{Aggr}(\vu_j^{(l)}, u_j \in \tilde{\mathcal{N}}(u_{i})) + \vu_i^{(l)} \right) \right), \label{eq:sage} \\
    \vu_i & = \operatorname{Comb} \left( { \vu_i^{(l)}: l \in { 0, 1, \dots, L } } \right), \label{eq:jknet}
\end{align}
where we set the learnable weights $B^{(l)}$ and $e_{ij}$ in the $\operatorname{Aggr}$ operation of Eq.~(\ref{eq:gnn}) as constant in Eq.~(\ref{eq:sage}) to reduce the parameter complexity. The $\operatorname{Comb}$ step in Eq.~(\ref{eq:jknet}) performs as a "skip connection" between different layers and usually does not involve additional parameters. Eq.~(\ref{eq:sage}) and Eq.~(\ref{eq:jknet}) are similarly defined for items.

We then conduct a two-layer GraphSAGE with JK-Net to generate the aggregated node representations $\vu$ and $\vt$ for all nodes from the source and target domains.
It is worth noting that it is also straightforward to employ other simpler GNN models such as LightGCN~\cite{he2020lightgcn} as the encoder as shown in Section~\ref{sec:gcn-arch}.

\xhdr{View generator}
Similar to the recent simplicity-driven design on the graph contrastive learning for recommendation~\cite{yu2022graph}, we choose to directly utilize the graph encoders from source domain $f^{s}$ and target domain $f^{t}$ as the natural contrasts for view generator. In this way, we could ease the burden of additional computation for graph augmentation types of contrastive that are used in~\cite{xie2022contrastive}. 

After obtaining the view of users ($\vu^{s}, \vu^{t}$) and items from ($\vt^{s}, \vt^{t}$) for source and target domains, we employ the separated paradigm by firstly conducting intra-CL to learn the relations within each domain then performing inter-CL to transfer the knowledge from source domain to target domain. The specific design of intra-CL and inter-CL is motivated from~\cite{xie2022contrastive}, which could effectively tackle the challenge of data sparsity, popularity bias, and diversity of CDR in matching.

\xhdr{Intra-CL loss}
In the first intra-CL stage, we only consider the neighbor-similarity-based loss to fully learn the self-supervised information from the sparse user-item interaction.
As in Figure~\ref{fig:framework}, this intra-CL loss projects all nodes into a common latent space, where the nodes are similar to their neighbors. To further utilize all edges as unsupervised information to facilitate the training process, in addition to user-item interactions, we propose to use BCELoss~\cite{gutmann2010noise} instead of the commonly used InfoNCE~\cite{oord2018representation} to implement this neighbor-similarity based intra-CL loss. In this way, the true label is the link (1 for existing, 0 for non-existing) between a node pair. Therefore, given a node $v^{s}_{i}$ ($v^{s}$ could be either a user $u^{s}$ or an item $t^{s}$) in the source domain, the intra-CL loss is formulated as follows:
\begin{align}
\mathcal{L}_{\text{intra}_\text{s}}(\vv^{s}) = -\sum_{v^{s}_{i}}\sum_{v^{s}_{j} \in \gN(v^{s}_{i})}\sum_{v^{s}_{k} \notin \gN(v^{s}_{i})} & ( \log(\sigma({\vv^{s\top}_i}\vv^{s}_j)) \notag \\
& + \log(1 - \sigma({\vv^{s\top}_i}\vv^{s}_k)) ),
\label{eq.L_intra_s}
\end{align}
where $v^{s}_{j} \in \gN(\vv^{s}_{i})$ is a sampled neighbor of $v^{s}_{i}$. $v^{s}_{k} \notin \gN(v^{s}_{i})$ is a randomly selected negative sample of $v^{s}_{i}$ ($v^{s}_{i}$ and $v^{s}_{k}$ are not connected).

Similarly, the intra-CL loss for a node $v^{t}_{i}$ in the target domain is:
\begin{align}
\mathcal{L}_{\text{intra}_\text{t}}(\vv^{t}) = -\sum_{v^{t}_{i}}\sum_{v^{t}_{j} \in \gN(v^{t}_{i})}\sum_{v^{t}_{k} \notin \gN(v^{t}_{i})}  & ( \log(\sigma({\vv^{t\top}_i}\vv^{t}_j)) \notag \\
& + \log(1 - \sigma({\vv^{t\top}_i}\vv^{t}_k)) ).
\label{eq.L_intra_t}
\end{align}

While InfoNCE is a more popular form of contrastive loss, the InfoNCE~\cite{oord2018representation} originates from BCELoss~\cite{gutmann2010noise} and BCELoss has been used to achieve contrastive learning especially for single domain graph learning~\cite{wu2018unsupervised,ren2021learning,zhao2022learning,chang2023knowledge}.  Using BCELoss also brings additional benefits for reducing the computational cost.

\xhdr{Inter-CL loss}
In the second inter-CL stage, we employ two types of contrastive losses: aligned user-based inter-CL and neighbor-similarity-based inter-CL as shown in Figure~\ref{fig:framework}.

\textit{Aligned user based inter-CL:} 
The majority of existing CDR methods \cite{man2017cross} adopt aligned users as their predominant mapping seeds across domains. We adhere to this idea and perform an aligned user-based inter-CL task. Each aligned user $u_i$ possesses two user representations $\vu^s_i$ and $\vu^t_i$ in the source and target domains, which are learned by two graph encoders during the intra-CL stage. Even though users may exhibit diverse preferences and behavior patterns in two domains, it is reasonable to assume that the source-domain representation $\vu^s_i$ should be more similar to its target-domain counterpart $\vu^t_i$ than any other representations $\vv^t_j$.

We define the user-based inter-CL loss $L_{\text{inter}_\text{u}}$ by the InfoNCE~\cite{oord2018representation} loss as follows:
\begin{equation}
\mathcal{L}_{\text{inter}_\text{u}}(\vu^{s}, \vv^{t}) = - \sum_{u_i} \log \frac{\exp (\mathrm{sim} (\vu^s_i,\vu^t_i)/\tau)}
{\sum_{v^{t}_i \in \gS^{t}_{u_i}} \exp (\mathrm{sim} (\vu^s_i,\vv^t_i)/\tau)},
\label{eq.L_inter_u}
\end{equation}
where $\tau$ is the temperature. $\gS^{t}_{u_i}$ is the sampled negative set collected from all other users/items in the target domain except $u_i$. The function $\mathrm{sim}(\cdot,\cdot)$ measures the similarity between a pair of embeddings, which is calculated with their cosine similarity.
Note that we do not use all examples in $\gS^{t}_{u_i}$ as negative samples for efficiency.

\textit{Neighbor-similarity based inter-CL:} 
In addition to the explicit alignments of users across domains, some indirect relations lack explicit mapping. Our goal here is to introduce more implicit cross-domain knowledge transfer paths between unaligned nodes in two domains. Following~\cite{xie2022contrastive}, we hypothesize that similar nodes in different domains should have similar neighbors (e.g., similar items may have similar users). Therefore, we adopt a neighbor-based inter-CL, which establishes indirect (multi-hop) connections between objects in different domains. 
The neighbor-based inter-CL loss $L_{\text{inter}_\text{n}}$ is then formulated with the aligned users $u_i$ and $u_i$'s neighbor set $\gN^t(u_i)$ in the target domain as follows:
\begin{equation}
\mathcal{L}_{\text{inter}_\text{n}}(\vu^{s}, \vv^{t}) = - \sum_{u_i} \sum_{v^{t}_i \in \gN^t(u_i)} \log \frac{\exp (\mathrm{sim} (\vu^s_i,\vv^t_i)/\tau)}
{\sum_{v^{t}_j \notin \gN^t(u_i)} \exp (\mathrm{sim} (\vu^s_i,\vv^t_j)/\tau)}.
\label{eq.L_inter_n}
\end{equation}
In $\mathcal{L}_{\text{inter}_\text{n}}$, the neighbor's representation $\vv^t_i$ in the target domain of an aligned user's representation $\vu^s_i$ in the source domain is the positive instance, while other target-domain representations $\vv^t_j$ are negative. This is justified by the assumption that related objects are connected and similar within a user-item graph, as enforced by the neighbor-similarity-based intra-CL loss in Eq.~(\ref{eq.L_intra_s}) and Eq.~(\ref{eq.L_intra_t}). The current positive samples $v^{t}_i \in \gN^t(u_i)$ can be extended to the neighbors from a multi-hop $r$-ego graph for improved generalization and diversity in CDR.
The neighbor-similarity-based inter-CL also increases the diversified knowledge transfer paths between two domains, especially for the cold-start items.

\begin{table*}[htbp]
    \centering
    \caption{Results of matching-related metrics on top-four Amazon datasets. All improvements of \method over baselines are significant (t-test with $p < 0.05$) and marked as bold.}
    \vspace{-2mm}
    \centering
    \label{tab:main_results_1}
    \begin{tabular}{l||c|c||c|c||c|c||c|c}
        \toprule
        \multirow{2}{*}{Model} & \multicolumn{2}{c||}{Books-Videos} & \multicolumn{2}{c||}{Books-Music} & \multicolumn{2}{c||}{Books-Elec} & \multicolumn{2}{c}{Books-Toys}  \\
        \cmidrule{2-9}
        & HIT@50 & HIT@100 & HIT@50 & HIT@100 & HIT@50 & HIT@100 & HIT@50 & HIT@100  \\
        \midrule
        MV-DNN~(\citeauthor{elkahky2015multi} \citeyear{elkahky2015multi}) & 0.1310  & 0.1848   & 0.1895   & 0.2074  & 0.0664 &  0.0958    & 0.0839     & 0.1487   \\
        EMCDR~(\citeauthor{man2017cross} \citeyear{man2017cross}) & 0.1310  &  0.1880  & 0.1632   & 0.2211  & 0.0646 &  0.0968    & 0.0911     & 0.1511   \\
        DTCDR~(\citeauthor{zhu2019dtcdr} \citeyear{zhu2019dtcdr}) & 0.0929  & 0.1648   & 0.1105   & 0.2124  & 0.0448 &  0.0682    & 0.0312     & 0.0767   \\
        \midrule
        HeroGraph~(\citeauthor{cui2020herograph} \citeyear{cui2020herograph}) & 0.1406  & 0.2068   & 0.1227   & 0.2301  & 0.0643 &  0.0948    & 0.1062     & 0.1557   \\
        GraphDR+~(\citeauthor{xie2021improving} \citeyear{xie2021improving}) & 0.1563  & 0.2431   & 0.1105   & 0.2421  & 0.0651 &  0.1245    &  0.1264    &  0.1828  \\
        CCDR~(\citeauthor{xie2022contrastive} \citeyear{xie2022contrastive}) & 0.1588  & 0.2545   &  0.1368  & 0.2737  &  0.0541 & 0.1048     & 0.1407     & 0.2350   \\
        COAST~(\citeauthor{zhao2023cross} \citeyear{zhao2023cross}) &  0.1475 & 0.2136   & 0.1265   & 0.2423  & 0.0683 &  0.1174    &  0.1117    & 0.1924   \\
        \method (ours) & \best{0.1616}   & \best{0.2613}   & \best{0.1435}   & \best{0.2847}  &  \best{0.0779}  &  \best{0.1238}    &  \best{0.1583}    &  \best{0.2490}  \\
        \bottomrule
    \end{tabular}
\end{table*}

\begin{table*}[htbp]
    \centering
    \caption{Results of matching-related metrics on bottom-four Amazon datasets. All improvements of \method are significant (t-test with $p < 0.05$) and marked as bold.}
    \vspace{-2mm}
    \centering
    \label{tab:main_results_2}
    \begin{tabular}{l||c|c||c|c||c|c||c|c}
        \toprule
        \multirow{2}{*}{Model} & \multicolumn{2}{c||}{Videos-Music} & \multicolumn{2}{c||}{Cloth-Music} & \multicolumn{2}{c||}{Kitchen-Music} & \multicolumn{2}{c}{Elec-Cloth }  \\
        \cmidrule{2-9}
        & HIT@50 & HIT@100 & HIT@50 & HIT@100 & HIT@50 & HIT@100 & HIT@50 & HIT@100  \\
        \midrule
        MV-DNN~(\citeauthor{elkahky2015multi} \citeyear{elkahky2015multi}) & 0.1205  & 0.1854   & 0.0867   & 0.1367  & 0.1123 & 0.1452     & 0.0170     & 0.0291   \\
        EMCDR~(\citeauthor{man2017cross} \citeyear{man2017cross}) & 0.1250  & 0.1786   & 0.1083   & 0.1550  & 0.1193 &  0.1432    & 0.0180     & 0.0313   \\
        DTCDR~(\citeauthor{zhu2019dtcdr} \citeyear{zhu2019dtcdr}) & 0.0536  & 0.0804   & 0.1350   & 0.1617  & 0.1250 &  0.1484    & 0.0115     & 0.0201   \\
        \midrule
        HeroGraph~(\citeauthor{cui2020herograph} \citeyear{cui2020herograph}) & 0.1182  & 0.1562   & 0.1567   & 0.1867  & 0.1477 &  0.1677    &  0.0088    & 0.0135   \\
        GraphDR+~(\citeauthor{xie2021improving} \citeyear{xie2021improving}) & 0.0759  & 0.1429   &  0.1533  & 0.1893  & 0.1494 &  0.1719    & 0.0198     & 0.0284   \\
        CCDR~(\citeauthor{xie2022contrastive} \citeyear{xie2022contrastive}) &  0.1356 & 0.1843   & 0.1660   & 0.2017  & 0.1538 &  0.1867    &  0.0220    & 0.0335   \\
        COAST~(\citeauthor{zhao2023cross} \citeyear{zhao2023cross}) &  0.1322 & 0.1792   & 0.1638   & 0.2075  & 0.1476 &  0.1786    &  0.0235    & 0.0379   \\
        \method (ours) &  \best{0.1484}  &  \best{0.1998}  & \best{0.1750}   & \best{0.2160}   &  \best{0.1619}  &  \best{0.1968}    &  \best{0.0298}     &  \best{0.0515}   \\
        \bottomrule
    \end{tabular}
\end{table*}

\subsection{Curriculum Scheduler in Inter-CL}
Considering the negative samples play an important role in the optimization of the inter-CL stage, we propose using a curriculum scheduler\cite{bengio2009curriculum,wang2021curriculum,chen2021curriculum,bian2021curriculum,wu2023curriculum} to utilize these negative samples effectively. We design an easy-to-hard curriculum training strategy to alleviate the impact of the potential noise hidden in the negative samples.

\xhdr{Difficulty measurer} In order to reflect the difficulty of the negative samples for GCL, we
propose to use graph complexity formalisms as difficulty criteria.
Inspired by~\cite{vakil2023curriculum}, we choose Katz centrality~\cite{katz1953new,newman2018networks} as the criterion of the difficulty measurer, since GNNs are trained through neural message passing in sampled subgraphs $\tilde{\gG^{s}}$ and $\tilde{\gG^{t}}$ for both the source and the target domains.

Specifically, the centrality of a node is determined by the centrality of its neighbors. Katz centrality quantifies the relative influence of a node within a network by counting the number of immediate neighbors and the number of walks between node pairs. The formula for the Katz centrality is as follows:
\begin{equation}\label{eq:katz}
    x_i = \alpha \sum_{j} \mA_{ij} x_j + \beta,
\end{equation}
where $\mA$ is the adjacency matrix and $\mA_{ij} = 1$ if $i$ and $j$ is connected. 
$x_i$ is the Katz centrality of node $i$, $\alpha$ is the attenuation factor, and $\beta$ controls the initial centrality.

Suppose that we first sample $N_{\text{neg}}$ negative samples for each node $u_{i}$,
for a positive-negative sample pair $(u_{i}, v_{j})$, we measure the difficulty as the sum of Katz centrality of them $d_{u_{i}, v_{j}} = d_{u_{i}} + x_{v_{j}}$. Then we reorder all $v_{j}$ in the negative sample set $\gS_{u_i}$ in descending order with respect to the difficulty score, which indicates that the higher Katz centrality corresponds to lower difficulty scores. This aligns with the intuition that the patterns of short-head items in the long-tail distribution can be more easily captured.
Note that the Katz centrality score for all nodes is pre-computed, therefore this process adds no additional computational complexity for CDR.

\xhdr{Curriculum scheduler}
With the ordered difficulty score for each pair of positive-negative samples, we can schedule the training of inter-CL in a curriculum manner.
For training inter-CL stage for $N_{\text{epoch}}$ epochs, we first extract the top 50\% easiest negative samples to form the negative set, \ie, $v^{t}_i \in \gS^{t}_{u_i}$ in Eq.~(\ref{eq.L_inter_u}) and $v^{t}_j \notin \gN^t(u_i)$ in Eq.~(\ref{eq.L_inter_n}), to optimize the model. Then for $N_{\text{step}}$ steps, we gradually select one negative sample with higher difficulty and include it into the negative set for curriculum contrasts. $N_{\text{step}}$ is defined as:
\begin{equation}\label{eq:scheduler}
    N_{\text{step}} = \frac{N_{\text{epoch}}}{N_{\text{neg}} / 2}.
\end{equation}
Although we choose this relatively static version of the implementation of the curriculum scheduler, it is worth mentioning that there are also other alternatives to achieving the curriculum scheduler, which we leave for future work.

\subsection{Multi-task Optimization}
\label{sec:multi_task_optimization}
Following other CDR models~\cite{xie2022contrastive}, we also conduct a multi-task optimization approach that combines the neighbor-similarity based intra-CL losses $\mathcal{L}_{\text{intra}_\text{s}}$ and $\mathcal{L}_{\text{intra}_\text{t}}$, the aligned user based inter-CL loss $\mathcal{L}_{\text{inter}_\text{u}}$, and the neighbor-similarity based inter-CL loss $\mathcal{L}_{\text{inter}_\text{n}}$ as follows:
\begin{align}
\text{Intra-CL Stage}:\text{ }\; \mathcal{L}_{\text{intra}} & = \lambda_{\text{intra}} \big( \mathcal{L}_{\text{intra}_\text{s}}(\vv^{s}) + \mathcal{L}_{\text{intra}_\text{t}}(\vv^{t}) \big), \label{eq:L_intra}\\
\text{Inter-CL Stage}:\text{ }\; \mathcal{L}_{\text{inter}} & = \lambda_{\text{inter}} \big( \mathcal{L}_{\text{inter}_\text{u}}(\texttt{stopgrad}(\vu^{s}), \vv^{t}) \notag \\
& + \mathcal{L}_{\text{inter}_\text{n}}(\texttt{stopgrad}(\vu^{s}), \vv^{t}) \big),
\label{eq:L_inter}
\end{align}
where $\lambda_{\text{intra}}$ and $\lambda_{\text{inter}}$ are loss weights for intra-loss and inter-loss, respectively.
$\lambda_{\text{intra}}$ and $\lambda_{\text{inter}}$ are shared between the source and the target domains and we did a grid search to set $\lambda_{\text{intra}}$ as $1.0$ and $\lambda_{\text{inter}}$ as $0.5$ across all datasets.

\begin{table}[htbp]
    \centering
    \caption{Results of matching-related metrics on our Industrial: Music2Videos dataset. All improvements of \method are significant (t-test with $p < 0.05$) and marked as bold.}
    \centering
    \label{tab:main_results_3}
    \begin{tabular}{l||c|c|c|c}
        \toprule
        Industrial: & \multirow{2}{*}{HIT@10} & \multirow{2}{*}{HIT@20} & \multirow{2}{*}{HIT@50} & \multirow{2}{*}{HIT@100} \\
        Music2Videos & & & & \\
        \midrule
        MV-DNN & 0.0109  & 0.0143   & 0.0262 & 0.0266 \\
        EMCDR  & 0.0308 & 0.0519 & 0.0433 & 0.0706 \\
        DTCDR & 0.0180  & 0.0247   & 0.0335 & 0.0366 \\
        \midrule
        HeroGraph & 0.0319  & 0.0602   &  0.0986  & 0.1374 \\
        GraphDR+ & 0.0351 & 0.0567   & 0.1012   & 0.1418 \\
        CCDR & 0.0819 & 0.1080 & 0.1489 & 0.1791 \\
        COAST & 0.0712 & 0.0946 & 0.1213 & 0.1582 \\
        \method (ours) &  \best{0.0852} & \best{0.1146} & \best{0.1547} & \best{0.1836} \\
        \bottomrule
    \end{tabular}
\end{table}

\begin{table*}[htbp]
    \centering
    \caption{Ablation study on the effectiveness of each module in \method on top-four Amazon datasets.}
    \vspace{-4mm}
    \centering
    \label{tab:ablation}
    \begin{tabular}{l||c|c||c|c||c|c||c|c}
        \toprule
        \multirow{2}{*}{Model} & \multicolumn{2}{c||}{Books-Videos} & \multicolumn{2}{c||}{Books-Music} & \multicolumn{2}{c||}{Books-Elec} & \multicolumn{2}{c}{Books-Toys}  \\
        \cmidrule{2-9}
        & HIT@50 & HIT@100 & HIT@50 & HIT@100 & HIT@50 & HIT@100 & HIT@50 & HIT@100  \\
        \midrule
        \method\#\# & 0.1482  & 0.2271   & 0.1199   & 0.2127  & 0.0598 &  0.1088    &  0.1274    &  0.1828  \\
        \method\# & 0.1570  & 0.2498   &  0.1393  & 0.2601  &  0.0663 & 0.1190     & 0.1397     & 0.2350   \\
        \method (ours) & \best{0.1616}   & \best{0.2613}   & \best{0.1435}   & \best{0.2847}  &  \best{0.0779}  &  \best{0.1238}    &  \best{0.1583}    &  \best{0.2490}  \\
        \bottomrule
    \end{tabular}
\end{table*}

\begin{figure}[htbp]
	\centering
		\includegraphics[width=0.48\columnwidth]{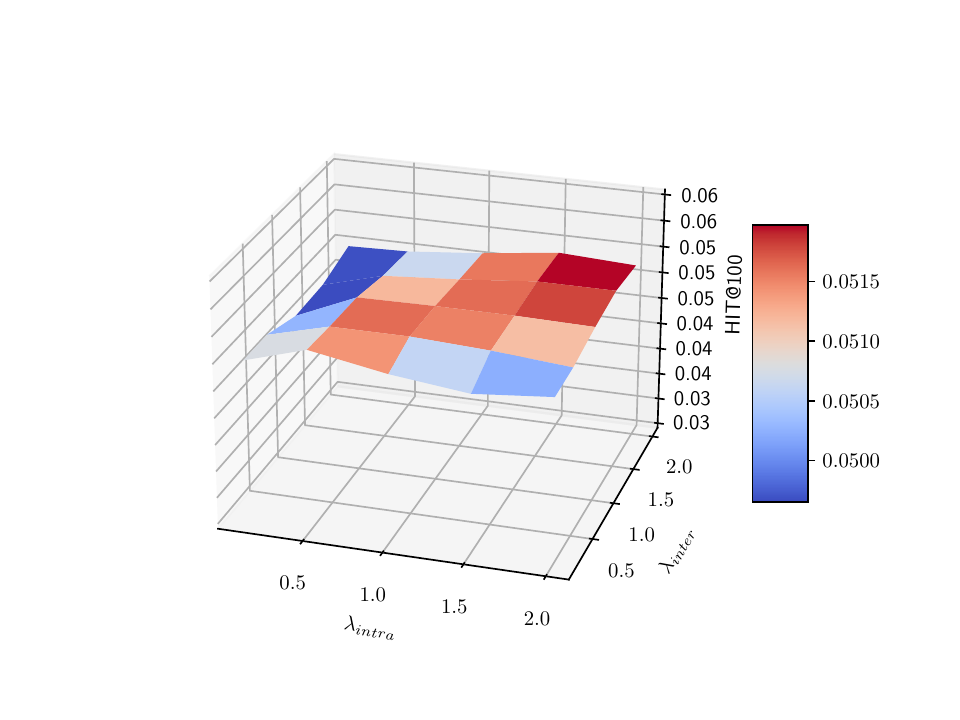}
		\includegraphics[width=0.48\columnwidth]{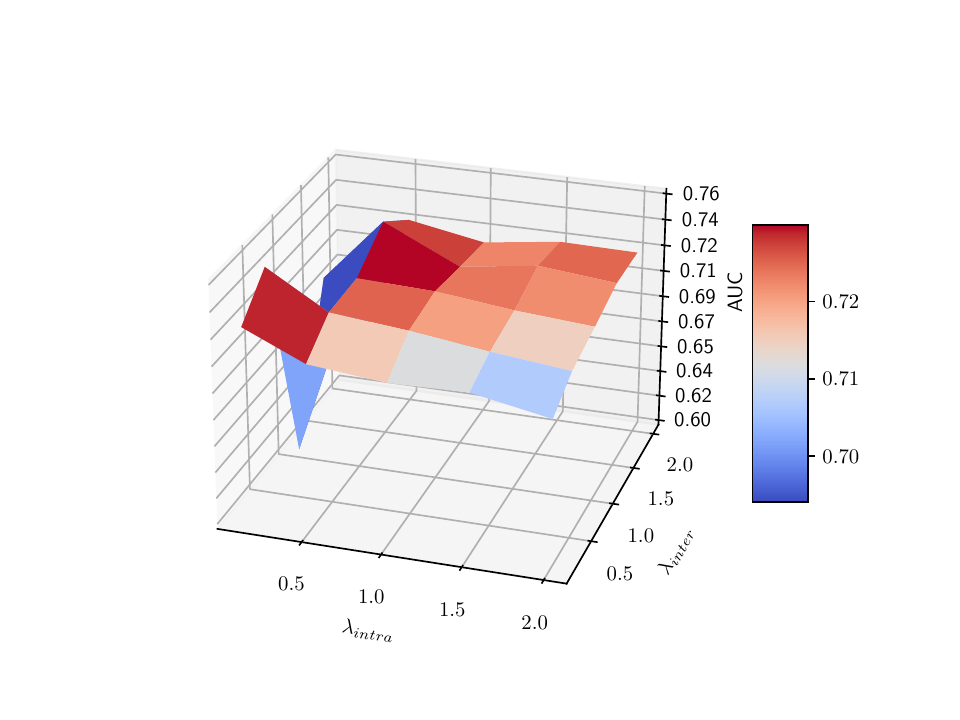}
        \vspace{-3mm}
        \caption{Sensitivity analysis on the loss weights $\lambda_{\text{intra}}$ and $\lambda_{\text{inter}}$ on Elec-Cloth dataset regarding both HIT@100 and AUC metrics. \label{fig:lambdas}}
        \Description{Sensitivity analysis on the loss weights $\lambda_{\text{intra}}$ and $\lambda_{\text{inter}}$ on Elec-Cloth dataset regarding both HIT@100 and AUC metrics. \label{fig:lambdas}}
\end{figure}

\section{Experiments}\label{sec:exps}
In this section, we evaluate \method on various real-world datasets and provide the ablations to demonstrate its effectiveness. 
Due to the page limits, the implementation details of our \method, the evaluation metric, and a description of the competing methods can be referred to in the Appendix.

\subsection{Experimental Setup}
\xhdr{Amazon datasets} We mainly use the Amazon datasets~\cite{he2016ups}, which contain users' behaviors in different domains.  We choose 7 domains from the Amazon datasets, which are Books, Movies\&TV (Videos), Digital Music (Music), Cloth, Electronics (Elec), Home\&Kitchen (Kitchen), and Toys, to create 8 source-target cross-domain scenarios, such as Books-Videos and Elec-Cloth. These scenarios cover both highly-correlated and less-correlated domains. We filter the behaviors of the source and target domains by the common users between domains. Table~\ref{tab:datastatistics} shows the detailed statistics of the filtered Amazon datasets\footnote{http://snap.stanford.edu/data/amazon}.

The target domain behavior numbers vary across different scenarios, for example, Books-Videos has 229,147 and Cloth-Music has 2,724. We follow the same dataset split for the target domain as~\cite{hu2018conet,ouyang2020minet,chen2023cross}, where the test set consists of the last behavior of each user, the validation set consists of the second to last behavior, and the training set consists of the rest behaviors. The feature for each user is its ID, and the features for each item are its ID and category.

\xhdr{Proprietary industrial datasets} To align the evaluation closer to real scenarios, we collect a new CDR dataset named Industrial: Music2Videos extracted from a real-world recommender system operated in proprietary Music and Video apps. We treat Music as the source domain and use its informative knowledge to help make recommendations for the target Video domain.
All data are pre-processed via data masking to protect the user’s privacy.
The users and their behaviors are randomly selected within a week and the detailed statistics can also be found in Table~\ref{tab:datastatistics}.
We split these behaviors into the training/validation/test sets similar to the strategy for Amazon datasets.

\subsection{Overall Performance Comparison}\label{exp:overall}
The overall recommendation performance in the task of matching in CDR of different models on Amazon datasets is presented in Table~\ref{tab:main_results_1}, Table~\ref{tab:main_results_2}, and Table~\ref{tab:main_results_3}. We can have the following observations:
\begin{itemize}
    \item \textbf{Graph-based matching method tends to have better performance than classical matching method.} This is attributed to the representation power of GNNs in structure learning, which effectively captures sparse interactions in the user-item graphs. Meanwhile, even the graph encoders are not implemented with fancy complicated GNN backbones, they have already achieved significant improvement by learning heterogeneous node embeddings are in the same semantic space, 
    \item \textbf{Our proposed \method brings consistent improvements over baselines.} Whether on open-source Amazon datasets or our proprietary industrial datasets, \method brings further improvement over either classical or graph-based matching methods. In particular, our \method achieves more than 2\% HIT@100 improvement over CCDR by sacrificing the additional graph augmentation intra-loss on more than half (Books-Music, Books-Elec, Books-Toys, Videos-Music,  Cloth-Music, Kitchen-Cloth, and Industrial: Music2Videos) of the settings. 
    \item \textbf{\method exhibits greater potential in dealing with less irrelevant source-target transfer and cold-start scenarios.} \method obtains about 0.1\% HIT@100 improvement in the Books-Videos scenario, where the source and target domains are quite similar, resulting in small differences in the optimal weights for different behaviors. However, our method shows more significant improvement in the Books-Elec scenario, where only specific book categories have an intuitive influence on electronic recommendation. Moreover, our method shows consistently larger HIT@100 enhancement (more than 2\% HIT@100) in the Books-Music, Videos-Music, Cloth-Music, Kitchen-Cloth, and Industrial: Music2Videos settings, where the target behaviors are much scarcer than the source domain, indicating the ability of our method to address the cold-start problem.
\end{itemize}

\subsection{Ablation Study}\label{sec:ablation}
To further examine the effectiveness of the two main designs of \method, we compare \method with its two ablation versions. We first remove the curriculum scheduler from \method to construct an ablation version \method \#, then consecutively remove the stop-gradient operation to build another ablation version \method \#\#. As shown in Table~\ref{tab:ablation}, we can observe that: 1) The curriculum scheduler that measures the difficulty of negative samples could enhance the performance of \method by persisting the potential noise in the graph structure. 2) The stop-gradient operation could contribute even more to the performance improvement since it effectively helps amplify the influence from the source domain by restricting the excessive gradient.

Given both intra-CL and inter-CL have already been verified essential for the matching in CDR from CCDR~\cite{xie2022contrastive}, we choose to skip this ablation study to avoid redundancy.

\subsection{Sensitivity Analysis of Loss Weights}\label{sec:lambdas}
Here we investigate the impact of two main hyperparameters, loss weights $\lambda_{\text{intra}}$ and $\lambda_{\text{inter}}$, on the performance (both HIT@100 and AUC metrics) regarding the Elec-Cloth dataset. The results are shown in Figure~\ref{fig:lambdas}. We can find that: 1) The \method performance is relatively sensitive to small loss weights $\lambda_{\text{intra}}$ and $\lambda_{\text{inter}}$; 2) Generally, the performance of \method is robust to loss weights, corresponding to the relatively flat surface resulting from the larger $\lambda_{\text{intra}}$ and $\lambda_{\text{inter}}$; 3) \method shows similar trends for both HIT@100 and AUC metrics, which indicates that the performance of our model is robust to the choice of evaluation metrics.
This sensitivity analysis of loss weights suggests that the users should set both $\lambda_{\text{intra}}$ and $\lambda_{\text{inter}}$ slightly larger during practice.

\subsection{Comparison of GNN Architectures}\label{sec:gcn-arch}
We also explore how the choice of different GNN architectures as the graph encoder of \method could influence the matching performance and report the results in Figure~\ref{fig:gcn_arc}. Four popular GNNs are selected for evaluation, \ie, GAT~\cite{velickovic2018gat}, GCN~\cite{ICLR2017SemiGCN}, LightGCN~\cite{he2020lightgcn}, and GraphSAGE~\cite{Hamilton2017Inductive}. To ensure a fair comparison, we set the number of layers as $2$, the embedding dimension as $128$, and employ JK-Net~\cite{Xu2018Representation} with concatenation as $\operatorname{Comb}$ accordingly for all methods. We can find that GAT and GraphSAGE achieve better HIT@100 for both datasets than GCN and LightGCN, which might be because the sampling strategy during training makes the average neighborhood aggregation in GCN and LightGCN weak for our task.
We can also see that the performance of \method does not significantly rely on any specific choice of the GNN architecture.
Considering the higher computational complexity and the larger amount of parameters of GAT, we choose to use GraphSAGE as the graph encoder of \method across all datasets in experiments.

\begin{figure}[htbp]
	\centering
	\begin{minipage}{0.485\columnwidth}
		\centering
		\includegraphics[width=\textwidth]{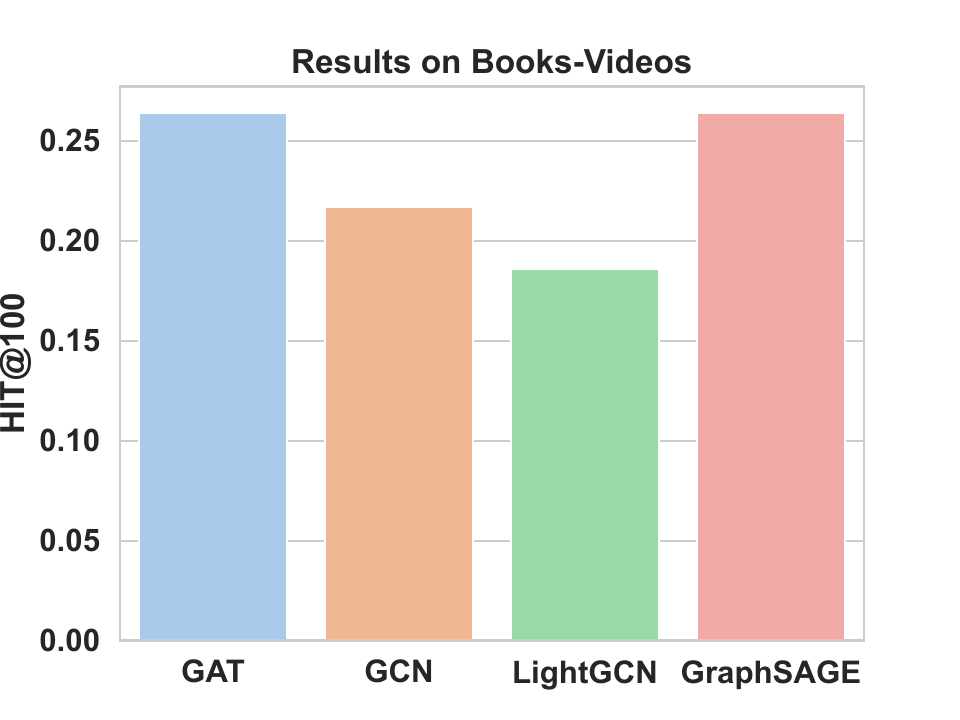}
		\subcaption{Books-Videos}\label{fig:gcn_arc_books2video}
	\end{minipage}
	\hfill
	\begin{minipage}{0.485\columnwidth}
		\centering
		\includegraphics[width=\textwidth]{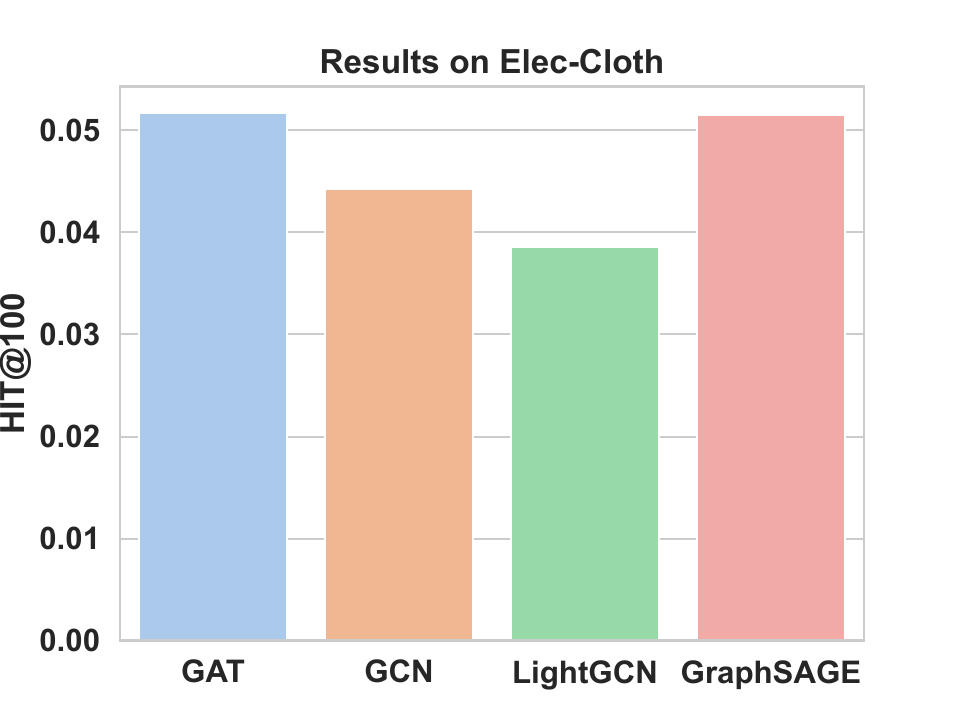}
		\subcaption{Elec-Cloth}\label{fig:gcn_arc_elec2cloth}
	\end{minipage}
	\vspace{-3mm}
        \caption{Performance comparison of different GNN architectures (GAT, GCN, LightGCN, and GraphSAGE). HIT@100 is reported as the evaluation metric here. \label{fig:gcn_arc}}
        \Description{Performance comparison of different GNN architectures (GAT, GCN, LightGCN, and GraphSAGE). HIT@100 is reported as the evaluation metric here. \label{fig:gcn_arc}}
\end{figure}

\subsection{Online Evaluation}
To further verify the effectiveness of \method in real-world scenarios, we conduct an online A/B test on our online recommender system within the Video App. The online baseline is GraphDR+ (target) which is trained solely on the target domain. We implement two baselines: GraphDR+ (source + target) and CCDR with the help of the source Music domain, and compare them with \method. We conduct an A/B test for 7 days. We focus on two online metrics in the target Video domain: i) CTR and ii) average user duration per capita (Duration). The results are reported in Table~\ref{tab:online_test}. We can find that all improvements of our \method over baselines regarding two online metrics are significant (t-test with $p < 0.05$), which further indicates the effectiveness of \method via online deployment.

\begin{table}[htbp]
    \centering
    \caption{Online A/B test results.}
    \vspace{-2mm}
    \centering
    \label{tab:online_test}
    \begin{tabular}{l||c|c}
        \toprule
        Model & CTR & Duration \\
        GraphDR+ (source + target) &  +0.76\% & +1.54\% \\
        CCDR &  +1.18\% & +2.14\% \\
        \method (ours) &  \textbf{+1.92\%} & \textbf{+3.65\%} \\
        \bottomrule
    \end{tabular}
\end{table}

\section{Conclusion}\label{sec:conclusion}
In this paper, we present a novel approach to cross-domain recommendation (CDR) that tackles the inherent challenges of applying contrastive learning methods across intra- and inter-domain tasks. Our curriculum scheduling framework (\method) explicitly separates the intra-domain contrastive learning (intra-CL) and inter-domain contrastive learning (inter-CL) tasks. We then apply an inter-domain curriculum scheduler to handle the complexity of cross-domain interactions. This approach stabilizes the training process and improves the quality of the user/item embeddings, resulting in better recommendation performance. Both the offline experiments and an online A/B test validate the effectiveness of SCCDR over a series of baselines, showing its state-of-the-art performance in CDR matching tasks. Our work highlights the importance of a nuanced approach to contrastive learning in cross-domain settings and opens up new avenues for further improving and optimizing CDR systems with curriculum learning strategies.

\bibliographystyle{ACM-Reference-Format}
\balance
\bibliography{refs}
\balance

\appendix

\section{Additional Experimental Setup}

\xhdr{Implementation details} 
All baselines are either directly adopted from their open-source codes or manually re-implemented based on their descriptions in original papers.
For \method,
we set the $\operatorname{Aggr}$ in GraphSAGE as the mean operation and the $\operatorname{Comb}$ in JK-Net as concatenation. The temperature $\tau$ for contrastive losses is set to $0.5$.
We use NetworkX~\cite{hagberg2008exploring} to calculate the Katz centrality for all nodes during the pre-processing step. We set $\alpha = 0.1$ and $\beta = 1.0$ as default in NetworkX since we observe the performance is not sensitive to them.
We set the loss weights $\lambda_{\text{intra}} = 1.0$ and $\lambda_{\text{inter}} = 0.5$ through grid search as analyzed in Section~\ref{sec:lambdas}.
For all methods, we set the dimension of output embedding as $128$, repeat every experiment $5$ times then report the mean performance. 
All embedding sizes are set as $64$, and the learning rate of these models is searched from $\{1\mathrm{e}^{-2},1\mathrm{e}^{-3},1\mathrm{e}^{-4}\}$ with weight decay as $5\mathrm{e}^{-4}$ and batch-size is searched from $\{1024,2048,4096\}$.

\xhdr{Evaluation metric and competing methods} 
We follow~\cite{xie2022contrastive} and other classical matching models \cite{xie2020internal,xie2021improving} to utilize the top $N$ hit rate (HIT@N) as our evaluation metric, which means that all models select top $N$ items from the overall corpora for each test instance. In other words, the full negative samples against each test instance are evaluated in HIT@N.
We should double clarify that \method focuses on CDR in matching, which cares whether good items are retrieved,
not the specific ranks that should be measured by ranking. Therefore, HIT@N is more suitable for matching than ranking metrics such as AUC and NDCG as indicated in CCDR~\cite{xie2022contrastive}. 
Even though, we also investigate the AUC metric in the sensitivity analysis of loss weights for complementary evaluation in Section~\ref{sec:lambdas}.

We implement several competitive baselines for comparisons, including classical and graph-based matching methods:
\textit{Classical matching methods:}
\begin{itemize}
 \item \textbf{MV-DNN}~(\citeauthor{elkahky2015multi} \citeyear{elkahky2015multi}) is a pioneer work that jointly learns from features of items from different domains and user features by introducing a multi-view DNN model.
 \item \textbf{EMCDR}~(\citeauthor{man2017cross} \citeyear{man2017cross}) is an embedding and mapping approach for matching tasks in CDR.
 \item \textbf{DTCDR}~(\citeauthor{zhu2019dtcdr} \citeyear{zhu2019dtcdr}) is based on Multi-Task Learning (MTL), and an adaptable embedding sharing strategy to combine embeddings of overlapped users.
\end{itemize}
\textit{Graph-based matching methods:}
\begin{itemize}
 \item \textbf{HeroGraph}~(\citeauthor{cui2020herograph} \citeyear{cui2020herograph}) proposes a heterogeneous graph framework for CDR and refines neighbor aggregation by recurrent attention.
 \item \textbf{GraphDR+}~(\citeauthor{xie2021improving} \citeyear{xie2021improving}) is an effective graph-based matching model that is directly constructed by modifying the single-domain version GraphDR on the joint network containing both source and target domains.
 \item \textbf{CCDR}~(\citeauthor{xie2022contrastive} \citeyear{xie2022contrastive}) is a novel framework to deal with CDR in matching by intra- and inter-domain CL with multi-task optimization, which is the most related work to \method.
 \item \textbf{COAST}~(\citeauthor{zhao2023cross} \citeyear{zhao2023cross}) aims to leverage rich content information and user interest alignment for bidirectional knowledge transfer in CDR.
\end{itemize}

\end{document}